\documentclass[11pt]{article}
\usepackage[hyperref]{acl}
\usepackage{times}
\usepackage{latexsym}
\usepackage{graphicx}
\usepackage{booktabs}
\usepackage{amsmath}
\usepackage{xcolor}
\usepackage{microtype}
\usepackage{inconsolata}
\usepackage{textcomp}
\newcommand{\head}{\bfseries}
\usepackage{tikz}
\usepackage{xurl} 
\usepackage[english]{babel}
\usepackage[autostyle, english = american]{csquotes}
\usetikzlibrary{positioning,arrows.meta,fit,backgrounds,shapes.geometric,decorations.pathreplacing,shapes.symbols,shadows,positioning,calc}
\MakeOuterQuote{"}

\definecolor{bubbleLeft}{RGB}{233,233,235}   
\definecolor{bubbleRight}{RGB}{0,132,255}    
\definecolor{textLeft}{RGB}{0,0,0}
\definecolor{textRight}{RGB}{255,255,255}

\newcommand{\leftx}{-0.4}
\newcommand{\rightx}{7.7}
\newcommand{\gap}{0.25cm}     

\newcommand{\code}[1]{\texttt{#1}}
\newcommand\satyrn{\textsc{Satyrn}}
\newcommand\tytan{\textsc{Tytan}}

\title{Tytan: Interactive Neurosymbolic Construction of\\Analytic Semantic Schemas from Relational Data}

\author{
    \quad Donna Hooshmand
    \quad Shubham Shahi
    \quad Cameron Barrie
    \quad Abhratanu Dutta\\
    \quad {\bf Marko Sterbentz}
    \quad {\bf Harper Pack}
    \quad {\bf Kristian J. Hammond}\\
    Northwestern University
}

\begin{document}
\maketitle
\begin{abstract}

From natural-language query interfaces to automated report generation, data analysis tools
need a description of the data: the real-world entities it contains, which columns function as measures or identifiers, and how tables connect into units of analysis. Today, this semantic layer is usually written by hand. This is a knowledge-acquisition bottleneck that limits the scalability of analytic systems, keeps non-technical users dependent on experts, and is itself error-prone.
We present \textsc{Tytan}, a system for automatically constructing an \emph{analytic semantic schema} from a relational database and, when available, a short user-provided description. \textsc{Tytan} combines symbolic analysis of the database with LLM-based semantic inference for entity proposal, role assignment, and naming. When the evidence leaves a decision ambiguous, \textsc{Tytan} asks the user a targeted natural-language question.

We evaluate \textsc{Tytan} on eight databases spanning real-world and benchmark domains along the three axes that define a schema's functional utility: (i) \emph{coverage}, are all important entities and features captured?; (ii) \emph{retrieval correctness}, do the schema's instructions actually reach the data?; and (iii) \emph{characterization accuracy}, are semantic types correct?

Across the seven reference domains, \textsc{Tytan} reaches every entity, attribute, and aggregable feature of the expert-corrected reference schemas (100\% coverage). Additionally, 100\% of its retrieval instructions execute correctly (1{,}678 of 1{,}678 self-generated claims), and semantic roles agree with the reference on 92-100\% of matched attributes. Checking the underlying data showed the small disagreement is in the reference, not in \textsc{Tytan}.
On a held-out blind test (a live, ten-table database with no declared keys), \textsc{Tytan}  recovers the full entity structure with verified keys and satisfies 100\% of the satisfiable expectations of five independent blind annotators.  

\end{abstract}
\vspace{-1.0em}

\section{Introduction}
Organizations often place a semantic layer between raw databases and the people or systems that use them. Instead of describing data only through tables, columns, and foreign keys, this layer represents the data in terms of real-world entities, attributes, and relationships \citep{chen1976entity,chen2012business}.
This intermediate layer is what allows a business-intelligence tool to interpret a request such as "average fire size per state" or a natural language interface to translate "which professors taught the most classes?" into appropriate joins and aggregations. Without this context, query systems would need to infer meaning directly from the schema, which is often ambiguous. As a result, LLM-based agents can misinterpret tables and relationships, and production natural-language-to-SQL systems remain unreliable on real-world databases that do not provide enough semantic information to resolve user intent \citep{floratou2024nl2sql, kim2020natural, li2024dawn}.

Despite its importance, this semantic layer is usually built by hand. Building a conceptual schema requires expertise in both the database's physical structure and the domain the data represents. This is the classic "knowledge-acquisition bottleneck" of ontology engineering \citep{mcginty2019knarm}. Additionally, this process is difficult to scale across the many datasets an organization may maintain, and manually written schemas can quickly fall out of date as the underlying database changes. It also leaves data consumers (who often lack database knowledge) dependent on more technical workers for routine data inquiry and analysis. 
This paper introduces \textsc{Tytan}, a system that is designed to address this challenge by automatically constructing a high-level semantic representation of a relational dataset with minimal manual effort. Given a relational database and an optional description of the dataset, \textsc{Tytan} produces an \emph{analytic semantic schema} we refer to as a \emph{ring}. This schema is a structured, self-constrained JSON artifact that identifies (i) the entities described by the data, (ii) each entity's attributes along with  an \emph{analytic role} (identifier, categorical, metric, or datetime). (iii) join paths connecting entities, (iv) linguistic surface forms (human-friendly names, descriptions grounded in real values, and label templates) that let downstream systems talk about the data and do analytical operations on the data. \emph{Rings} adopt the format of the \textsc{Satyrn} analytics platform \citep{sterbentz2024satyrn}, but the representation is system-agnostic. It's a semantic data catalog that any schema-aware tool can use.

\begin{figure*}[t]
\centering
\resizebox{\textwidth}{!}{
\begin{tikzpicture}[
  font=\small,
  node distance=5mm and 7mm,
  stage/.style={draw=black!60, rounded corners=2pt, fill=blue!8,
                align=center, text width=24mm, inner sep=3pt},
  io/.style={draw=black!60, rounded corners=2pt, fill=gray!12,
             align=center, inner sep=3pt},
  ring/.style={draw=black!70, thick, rounded corners=3pt, fill=orange!18,
               align=center, text width=25mm, inner sep=3pt},
  consumer/.style={draw=black!60, rounded corners=2pt, fill=green!12,
                   align=center, text width=22mm, minimum height=8mm,
                   inner sep=2pt},
  arr/.style={-{Stealth[length=2.2mm]}, thick, black!70},
  looparr/.style={-{Stealth[length=2mm]}, black!60, dashed}
]

\node[io, cylinder, shape border rotate=90, aspect=0.1,
      minimum width=14mm, minimum height=13mm, fill=gray!15] (db)
      {\scriptsize DB\\[-1pt]\scriptsize instance};
\node[io, below=4mm of db, text width=17mm, fill=yellow!20] (desc)
      {\tiny User input\\[-2pt]\tiny\emph{(optional)}};

\node[stage, right=9mm of db, yshift=-4mm] (profile)
      {\textbf{Structural}\\\textbf{profiling}\\[2pt]
       {\tiny\color{black!55}keys, samples,}\\[-2pt]
       {\tiny\color{black!55}cardinalities, nulls}};

\node[stage, right=of profile] (propose)
      {\textbf{LLM semantic}\\\textbf{inference}\\[2pt]
       {\tiny\color{black!55}entities, roles,}\\[-2pt]
       {\tiny\color{black!55}names, joins}};

\node[stage, right=of propose] (verify)
      {\textbf{Grounding \&}\\\textbf{verification}\\[2pt]
       {\tiny\color{black!55}value-overlap probes,}\\[-2pt]
       {\tiny\color{black!55}verified keys, guards}};

\node[stage, right=of verify] (clarify)
      {\textbf{Clarification}\\\textbf{loop}\\[2pt]
       {\tiny\color{black!55}targeted questions,}\\[-2pt]
       {\tiny\color{black!55}conservative defaults}};

\draw[looparr] (verify.north) to[bend right=25]
      node[above, font=\tiny, text=black!55] {iterative sweep: unclaimed tables/columns}
      (propose.north);

\node[io, above=7mm of clarify, text width=13mm, fill=yellow!20] (user)
      {\scriptsize User};
\draw[looparr] (clarify.north) to[bend left=15] (user.south west);
\draw[looparr] (user.south east) to[bend left=15] (clarify.70);

\node[ring, right=9mm of clarify] (ringbox)
      {\textbf{Ring}\\[2pt]
       {\tiny\begin{tabular}{@{}c@{}}
         semantic layer
       \end{tabular}}};

\node[consumer, right=8mm of ringbox, yshift=11mm] (satyrn)
      {\scriptsize analytics \& report\\[-2pt]\scriptsize generation (\textsc{Satyrn})};
\node[consumer, right=8mm of ringbox] (nlq)
      {\scriptsize natural-language\\[-2pt]\scriptsize query interfaces};

\node[consumer, right=8mm of ringbox, yshift=-11mm] (bi)
      {\scriptsize Business Intelligence (BI) tools /\\[-2pt]\scriptsize semantic layers};

\draw[arr] (db.east) -- (profile.west|-db.east);
\draw[arr] (desc.east) to[bend right=12] (propose.south west);
\draw[arr] (profile) -- (propose);
\draw[arr] (propose) -- (verify);
\draw[arr] (verify) -- (clarify);
\draw[arr] (clarify) -- (ringbox);
\draw[arr] (ringbox.east) to[bend left=10] (satyrn.west);
\draw[arr] (ringbox.east) -- (nlq.west);
\draw[arr] (ringbox.east) to[bend right=10] (bi.west);

\draw[looparr] (db.south) to[bend right=20]
      node[below, font=\tiny, text=black!55, pos=0.55]
      {every proposal validated against the instance}
      (verify.south);

\begin{scope}[on background layer]
  \node[draw=black!40, rounded corners=4pt, fill=blue!3, inner sep=4mm,
        fit=(profile)(propose)(verify)(clarify)(user),
        label={[font=\small\bfseries, text=black!60, anchor=north west,
                xshift=1.5mm, yshift=-1mm]north west:\textsc{Tytan}}] {};
\end{scope}
\end{tikzpicture}}
\caption{\textsc{TYTAN}{} overview. A database instance and an optional
one-sentence description enter the pipeline. Symbolic profiling, LLM
proposals, and deterministic verification interleave to produce a
\emph{ring}: a self-contained semantic catalog (an example of a ring is given in Figure~\ref{fig:ring}). Decisions the evidence
cannot settle go to the user as clarification questions.}
\label{fig:overview}
\end{figure*}
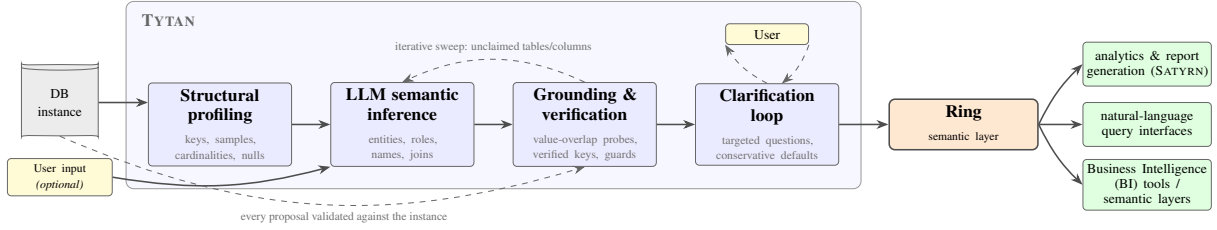

Limitations of symbolic and neural methods make the automatic construction of schemas challenging. Symbolic techniques, like key discovery and statistical type detection can recover structural properties of a database but can't reliably infer what those properties mean or the semantic importance of them. They may identify a column's data type or detect a candidate primary key, but they cannot determine that \code{household\_income} should be treated as an aggregable field while \code{zip\_code} is Categorical. They may also fail to recognize that a table containing student, course, and grade information represents a junction table rather than a standalone entity. LLMs are capable of making such semantic inferences, but they might hallucinate columns, invent joins between unrelated tables, and are not always consistent across multi-step construction. The central design on \textsc{Tytan} leverages the strengths of each method, while creating safeguards to eliminate the limitations. Every LLM proposal is validated against the database: name-and-type matching for column mappings, value-overlap for detecting joins, uniqueness for detecting keys, sample-value evidence for detecting types. Additionally, properties that can be computed directly from the database are handled through deterministic procedures. If the available evidence is still insufficient, \textsc{Tytan} asks the user a focused natural-language question and uses the response to update the schema (Figure~\ref{fig:overview}). This human-in-the-loop paradigm improves robustness and transparency, positioning \textsc{Tytan} as a general-purpose system for automated semantic modeling, unlike prior work that focuses narrowly on the type detection or static schema recovery.

We make the following contributions: 
\begin{itemize}
\item We define \emph{analytic schema construction} as the task of constructing a semantic representation from the database instance. This representation must support analytical query planning and validation, as well as natural-language descriptions of query results. 

\item We present \textsc{Tytan}, an interactive neurosymbolic pipeline for analytic schema construction. We describe how it validates LLM-generated inferences against the database by validating proposed joins with deterministic checks, using uniqueness for testing entity keys, and using sample values for refining type assignments without replacing semantic judgments. When the available evidence is not enough to resolve a decision, it triggers an interactive clarification to handle uncertainties. 

\item We define a functional evaluation framework that measures whether a generated schema can support the tasks it's meant to serve. The evaluation covers entity and feature coverage against expert-corrected references; the accuracy of its semantic type; and retrieval correctness,by validating claims that the schema makes against the source database. 

\item We evaluate \textsc{Tytan} on eight databases, including seven reference domains and one held-out test. We report results across all three evaluation dimensions and test the generated schemas in an end-to-end analytics platform. 
\end{itemize}

\section{Related Work}
\label{sec:related}

\paragraph{Conceptual modeling and schema recovery.}
Deriving conceptual models from other artifacts is a long-standing goal
\citep{chen1976entity, batini1986comparative, yu2006schema, kelloumenouer2022survey}. 
Earlier work generated entity-relationship diagrams from controlled natural-language requirements using syntactic patterns and hand-built lexicons \citep{chen1983english}. these methods worked within small, domain-specific datasets with limited vocabulary but did not generalize well beyond them. Some reverse-engineering tools such as
SchemaSpy\footnote{\url{https://schemaspy.org}} visualize the structure declared in a database schema, but their practicality heavily depends on primary- and foreign-key constraints. However, such constraints are often missing from real-world databases \citep{dohmen2024schemapile}.
Such tools surface what the schema already states; they cannot infer what it omits. A table's meaning, an attribute's role, an undeclared relationship, all remain out of reach.
More recent work uses LLMs to generate conceptual schemas from textual specifications \citep{prokop2024enhancing, divljan2025llm}. These methods operate on descriptions rather than instances, so they can't directly check whether the generated tables, attributes, or relationships are supported by the data. 
Other recent work infers conceptual models over heterogeneous tabular data from column names and sampled values \citep{wu2025conceptual}. This is the closest work to ours, and it does its job well: it recovers conceptual structure across many heterogeneous tables to help users find and understand relevant datasets. Our goal is different. We construct a schema a downstream system can compute over, not one for dataset discovery. That is why \textsc{Tytan} adds what analysis requires and their setting does not: entity keys, analytic roles, and column-level join paths checked against the data. The construction of the schema is also fully automated, with no way to incorporate user feedback, and the evaluation measures how closely the schema's structure matches a reference rather than what a user can do with it. 
\textsc{Tytan} is designed for a different use case and creates the complementary artifact: it constructs a schema that can be used directly by downstream analytic systems, checks its structural claims against the database, and asks the user to resolve decisions that remain ambiguous. Thus, we evaluate the schema in terms of the analytic tasks it can support, rather than structural similarity alone. 
(\S\ref{sec:eval}).

\paragraph{Semantic type detection and table understanding.}
A related body of work is focused on assigning semantic types to individual columns \citep{venetis2011recovering}. Sherlock learns column-type classifiers from large collections of tables \citep{hulsebos2019sherlock}. Sato extends this idea by using relationships between columns as part of the prediction process \citep{zhang2019sato}. TURL supports both column annotation and relation extraction \citep{deng2020turl}, while Doduo uses fine-tuned language models to identify column types \citep{suhara2022doduo}. More recent approaches use LLMs to label columns with little or no task-specific training \citep{korini2023chatgpt,feuer2024archetype}. This reflects the broader use of foundation models for data-wrangling tasks \citep{narayan2022wrangle, kayali2024chorus}.
Column type detection addresses an important part of schema construction, but only at the level of individual attributes, usually within a single table. It does not determine which tables represent entities or how those entities are connected. In \textsc{Tytan}, column typing is one step in a broader process. Type assignments are determined by the LLM's interpretation of the column, as well as deterministic evidence from the values stored in the database (\S\ref{sec:types}).

\paragraph{Key and join discovery.}
a long-standing database problem has been discovering keys and inclusion dependencies that are not explicitly declared in the schema \citep{rahm2001survey, rostin2009machine, papenbrock2015divide, jiang2020holistic}. In practice, exhaustive dependency discovery is both costly and prone to false positives. Two columns may appear related simply because they contain overlapping ranges of small integer identifiers, even when no meaningful relationship exists between them \citep{zhang2010multicolumn}. Rather than trying to recover every possible dependency, \textsc{Tytan}{} builds a conservative relationship graph intended for analytic use. Declared constraints are accepted directly. A join that is not declared is included only when it passes several checks, including value overlap, key-role and specificity tests, and an LLM-based plausibility assessment above a strict confidence threshold (\S\ref{sec:joins}).

\paragraph{Natural-language interfaces to data.}
Text-to-SQL research has produced increasingly capable systems and widely used benchmarks \citep{yu2018spider, li2023bird, fu2023catsql}. Even so, deploying these systems on real databases remains difficult. One recurring problem is that table definitions alone often do not provide enough context to resolve what a user means or how a query should be constructed \citep{floratou2024nl2sql, kim2020natural, li2024dawn}. In practice, organizations address this gap by adding semantic layers or other forms of manually written database documentation.\footnote{See, for example, IBM's semantic-layer guidance (\url{https://www.ibm.com/think/topics/semantic-layer}) and TigerData's proposal for self-describing databases (\url{https://www.tigerdata.com/blog/the-database-new-user-llms-need-a-different-database}).} \textsc{Tytan}{} focuses on creating this missing context. It derives a semantic schema from the database itself and asks for user input only when the available evidence does not resolve an ambiguity.

\paragraph{Analytics-augmented generation.}
\textsc{Satyrn} \citep{sterbentz2023lightweight, sterbentz2024satyrn} produces factual analytic reports by planning symbolic analyses over a manually annotated ring schema and passing the resulting facts to an LLM. Its published evaluation relied on expert-written rings for eight domains, illustrating the manual annotation burden that motivates this work. We use \textsc{Satyrn}{} to evaluate \textsc{Tytan} in a downstream setting (\S\ref{sec:eval-extrinsic}), although the schemas produced by \textsc{Tytan} are not tied to \textsc{Satyrn}.

\section{Problem: Analytic Schema Construction}
\label{sec:problem}

\subsection{What the schema must contain}
\label{sec:ring}
An analytic system needs a schema to have an analytical structure, which describes both how the data can be used; and a linguistic surface, which shows how it should be presented \citep{sterbentz2024satyrn}.  It must identify the main units of analysis, specify which operations are valid for each attribute, and record the join paths needed for questions that span multiple entities. It must also provide readable names and descriptions so that results can be expressed in natural language rather than with raw column names such as \code{patient\_id}.

\begin{figure}[t]
\centering
\resizebox{0.98\columnwidth}{!}{
\begin{tikzpicture}[
  font=\scriptsize,
  block/.style={draw=black!55, rounded corners=2pt, align=left,
                text width=54mm, inner sep=3pt},
  grouplbl/.style={font=\tiny\itshape, text=black!60, align=left,
                   text width=13mm},
  brace/.style={decorate, decoration={brace, amplitude=3pt}, black!60}
]

\node[block, fill=purple!8] (ds) {%
\textbf{data\_source}\\[1pt]
\ttfamily\tiny type: postgres\quad connection: <url>\\
\ttfamily\tiny joins: rent\_rate.region\_id \(\rightarrow\)\\
\ttfamily\tiny \hphantom{joins: }region\_information.region\_id\ \ (m:1)};

\node[block, fill=orange!10, below=2.5mm of ds] (ent) {%
\textbf{entity: RegionInformation}\ \ \tiny(table \texttt{region\_information})\\[1pt]
\ttfamily\tiny id: [region\_id]\ \ isa: integer\ \ \rmfamily\tiny\emph{(verified unique)}\\
\ttfamily\tiny reference: ``\{RegionName\}, \{StateName\}''};

\node[block, fill=blue!7, below=2.5mm of ent] (attr) {%
\textbf{attributes} \tiny(one of four, expanded)\\[1pt]
\ttfamily\tiny SizeRank:\\
\ttfamily\tiny\ \ source: \{table: region\_information,\\
\ttfamily\tiny\ \ \hphantom{source: \{}columns: [size\_rank]\}\\
\ttfamily\tiny\ \ isa: integer\quad type: [Numeric, Metric]\\
\ttfamily\tiny\ \ nicename: ``Size Ranking''\\
\ttfamily\tiny\ \ description: ``...rank by size. e.g.\ 1, 2, 3''};

\node[block, fill=green!8, below=2.5mm of attr] (agg) {%
\textbf{aggregable\_attributes} \tiny(reachable via joins)\\[1pt]
\ttfamily\tiny AverageRent \(\leftarrow\) rent\_rate.average\_rent\\
\ttfamily\tiny\ \ joins: [rent\_rate.region\_id \(\rightarrow\) ...]\\
\ttfamily\tiny\ \ type: [Numeric, Metric]\ \ \rmfamily\tiny\emph{``avg rent per region''}};

\node[block, fill=orange!10, below=2.5mm of agg] (ent2) {%
\textbf{entity: RentRate}\ \ \tiny(table \texttt{rent\_rate})\\[1pt]
\ttfamily\tiny id: [region\_id, date]\ \ isa: string\\
\rmfamily\tiny\emph{(no declared key --- composite verified against the data)}\quad
\ttfamily\tiny ...};

\draw[brace] ([xshift=1.5mm]ds.north east) -- ([xshift=1.5mm]ds.south east)
  node[grouplbl, midway, right=4pt] {retrieval\\instructions};
\draw[brace] ([xshift=1.5mm]ent.north east) -- ([xshift=1.5mm]ent.south east)
  node[grouplbl, midway, right=4pt] {unit of\\analysis\\+ label};
\draw[brace] ([xshift=1.5mm]attr.north east) -- ([xshift=1.5mm]attr.south east)
  node[grouplbl, midway, right=4pt] {analytic\\role +\\surface\\forms};
\draw[brace] ([xshift=1.5mm]agg.north east) -- ([xshift=1.5mm]agg.south east)
  node[grouplbl, midway, right=4pt] {cross-entity\\reachability};
\draw[brace] ([xshift=1.5mm]ent2.north east) -- ([xshift=1.5mm]ent2.south east)
  node[grouplbl, midway, right=4pt] {verified\\identity};
\end{tikzpicture}}
\caption{Anatomy of a generated ring (housing-rent database, abridged).
The right-hand annotations mark what each block is \emph{for}:
\texttt{data\_source.joins} and per-attribute \texttt{source} blocks are
executable retrieval instructions; each entity's verified \texttt{id} and
\texttt{reference} make it a unit of analysis with a human-readable
label; \texttt{isa}/\texttt{type} license analytic operations; and
aggregable attributes expose related entities' features through validated
join paths. \texttt{RentRate} illustrates the no-declared-key case: its composite id was verified unique against the data.}
\label{fig:ring}
\end{figure}
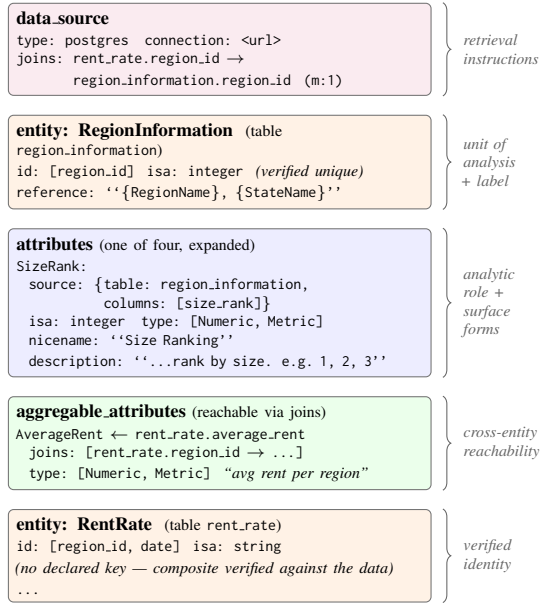

The ring format captures both (Figure~\ref{fig:ring}). A ring contains:
\begin{itemize}
  \item \textbf{Entities:} Real-world concept (e.g., \code{Wildfire}, \code{Patient}) that maps to a source table, and declares an \emph{id}: a column or column set that uniquely identifies different instances.
  \item \textbf{Attributes:} Each attribute carries provenance
  (\code{source} table and columns), a physical storage type (\code{isa} $\in$ \{string, integer, float, date, datetime, boolean\}), an analytic role (\code{type}): \emph{Categorical},
  \emph{Datetime}, or \emph{Numeric} -- plus an additional Metric tag for attributes that represent a measurement of the corresponding entity, a human-friendly name, and a description grounded in example values drawn from the data. The storage type and the analytic role are deliberately independent: an insurance ID stored as an integer  but would have a \emph{Categorical} type; averaging out a set of IDs produces a meaningless number so numerical operations should not be applied such attributes.
  
  \item \textbf{Relationships:} A global list of verified join paths, plus per-entity \emph{aggregable attributes}: attributes of related entities reachable through those joins, annotated with the path and its direction (many-to-one or one-to-many). On a many-to-one path the related attribute acts like one of the entity's own columns. On a one-to-many path, questions about it only make sense with aggregation.
  
  \item \textbf{Surface forms:} Entity and attribute names, and a
  \emph{reference template} per entity: a short label built from one or two identifying attributes (e.g., \code{''\{FirstName\} \{LastName\}''}) used when results mention individual instances.
\end{itemize}

What the ring \emph{excludes} is also deliberate. It records the available data, where it comes from, how records are identified, and which analytic operations each attribute supports. The downstream system is responsible for carrying out those operations. For example, a \code{match\_date} labeled as \emph{Datetime} can be grouped by year, month, or season even if the database contains no separate columns for those values. Counts, rankings, and standings are likewise results of analysis rather than attributes of the schema.
Including these derived values in the ring would duplicate work. Schema-aware system can perform these operations when needed and would tie the schema to the features of a particular user. This distinction also affects evaluation. A question is supported when the necessary data can be reached through the ring and the relevant operations are correctly specified, not only when the schema contains a column whose name closely matches the question (\S\ref{sec:eval-blind}).

\subsection{Problem statement}

Given a database instance $D$, including its tables, columns, declared constraints, and stored values, along with an optional natural-language description $d$, the goal is to construct a ring $R$ with three properties: (i) Semantically relevant tables and columns in $D$ should be represented or reachable in $R$ (ii) The contents of $R$ must agree with the database and should not be hallucinations: referenced tables,columns and proposed joins must exist in the database. (iii) $R$ should support as broad a range of valid analytic questions as possible for use of a downstream schema-aware system. The construction process may inspect the database directly and ask the user a small number of focused questions when the available evidence is not enough to prevent ambiguities. The second requirement is especially important. A schema may appear reasonable while still containing a join between unrelated tables or assigning a non-unique column as an entity key. These errors may not be obvious from the schema itself, but they can invalidate any downstream analysis that depends on them. We therefore treat such errors as primary failure cases in the evaluation (\S\ref{sec:eval-intrinsic}), and much of \textsc{Tytan}'s design is devoted to detecting and preventing them.

\section{The \textsc{Tytan} System}
\label{sec:system}
\textsc{Tytan} operates in three
interleaved phases: (i) \emph{symbolic
analysis} of the database, (ii) \emph{LLM-driven semantic inference}  over structural evidence to propose a conceptual schema, and (iii) \emph{interactive clarification} with the user to handle uncertainties.
This section follows the pipeline (Figure~\ref{fig:overview}) stage by stage. The pipeline starts with deterministic symbolic profiling; after that, every stage follows the same pattern: LLM proposes, and the deterministic rules verify.

\subsection{Structural profiling and Symbolic analysis}
\label{sec:profiling}
\textsc{Tytan} begins by collecting deterministic information from the database: its tables and columns, any declared primary or foreign keys, samples of distinct column values, cardinalities, and null rates. 
\textsc{Tytan} also accepts flat files such as CSV and Excel. These types of files are loaded into a relational store so that the same machinery applies but unlike a
managed database, there are no primary keys, no foreign keys, and often no reliable column types. In this setting the metadata sources above are simply empty, and profiling can only supply value-based evidence. Every key must be verified for
uniqueness against the data, every join must be discovered through
value overlap, and column types must be inferred from samples (\S\ref{sec:joins}, \S\ref{sec:types}). Our evaluation includes settings where no such metadata is available
(\S\ref{sec:eval}).

\subsection{Entity proposal and grounding}
\label{sec:entities}
Using an optional the database description and the extracted evidence, the LLM first proposes a small set of domain entities and their likely attributes. For example, it may infer that a dataset describes wildfires and that each wildfire has a location, cause, and size. The next stage connects these concepts to the database. Each proposed entity is matched to a source table, and each attribute is matched to one or more columns using column names, data types, and sample values.

The resulting mappings are then checked and cleaned up deterministically. When two proposed entities map to the same table, they are merged rather than allowed to overwrite one another. The entity name may also be revised when the source table makes the row-level meaning clearer. For example, a proposed entity named ``\texttt{Location}'' may be renamed \code{Wildfire} if the matched table contains one row per wildfire incident.

After the initial mappings are complete, \textsc{Tytan} examines the tables and columns that remain unassigned. Each unmapped table is placed into one of four categories. An \emph{entity} represents a real-world instance and the main unit of analysis. A \emph{link table} records a many-to-many relationship between entities. A \emph{passthrough} table is a small lookup or conversion table, such as a grade-conversion chart or status-code table, whose values should remain accessible through joins even though the table is not itself a unit of analysis. The \emph{ignore} category is reserved for technical tables such as migration logs. The classifier avoids using \emph{ignore} unless there is clear evidence that a table has no analytic value, reducing the chance that useful lookup data will be discarded. Finally, any unmapped columns in an already assigned table are added to the corresponding entity as attributes. This final pass helps ensure that the generated ring covers the database rather than only the fields selected during the initial proposal stage.

\subsection{Join discovery and validation}
\label{sec:joins}
When a foreign key is declared in the database, \textsc{Tytan} uses it directly. The difficult case is when the database does not contain any metadata or relationships as constraints. To recover these missing joins, \textsc{Tytan} combines two sources of candidates: relationships proposed by the LLM, especially when interpreting link tables, and relationships suggested by column names and overlapping values. Every undeclared candidate then passes through the following checks:

\begin{enumerate}
\item \textbf{Value-overlap.} A candidate join whose columns share no values cannot connect any rows and is removed. The check uses all distinct values rather than a sample, so zero overlap is enough to rule the join out. We apply this check even when the join was proposed confidently by the LLM.

\item \textbf{Deterministic key-role guards.} Several kinds of column
pairs are excluded before LLM verification. A source column that is itself a primary key, declared or inferred, is not treated as a foreign key, since overlap between two key columns often comes from unrelated small-integer ranges. Columns classified as Numeric measures are also excluded, because foreign keys act as identifiers, while counts such as goals or saves (in soccer) may happen to overlap with sequential IDs. A column that already has a verified reference is not assigned a second referent.
Finally, if a column's values overlap the keys of three or more different tables, all of its candidates are dropped: the overlap is
more likely to reflect a generic value range than a specific relationship. This count is computed over all overlapping targets before any other filter is applied, so the guard cannot be undercut by earlier skips.

\item \textbf{LLM verification.} After the checks above, the remaining candidate joins are checked for semantic plausibility using the two columns' names, types, and sample values. For example, the model must decide whether \code{Student.Major} $\rightarrow$ \code{Department.DNO} (department code) represents a real reference, or whether columns such as \code{physician.code} and \code{medication.code} merely contain values from overlapping ID ranges. A candidate is accepted only when the model assigns confidence above a fixed threshold. If verification fails, the join is dropped.

\end{enumerate}

The resulting graph contains declared foreign keys and undeclared joins that have passed both data-based and semantic checks. \textsc{Tytan} then searches this graph to determine which attributes of related entities are reachable from each entity. It distinguishes attributes reached through a direct many-to-one relationship from those reached through reverse or multi-hop paths. Related entity identifiers are exposed as countable targets, while unhelpful paths, such as a route back to the source entity's own key, are removed.
Each edge in the graph is a join and also keeps a canonical direction. This direction is used when the join is serialized rather than relying on the endpoint order returned by the graph library. 

Two additional steps address cases that would otherwise leave the join graph incomplete. First, entities introduced during the iterative coverage sweep are assigned an identifier that other tables can reference. If the LLM did not identify a key, \textsc{Tytan} uses declared metadata when available or selects an id-like column whose values are verified to be unique. Without this step, the entity would be absent from the key map and could not receive incoming relationships.

Second, \textsc{Tytan} preserves multiple foreign keys between the same pair of tables. Both \code{matches.home\_team\_id} and \code{matches.away\_team\_id}, for example, may reference \code{teams}, and \textsc{Tytan} keeps them as separate joins rather than letting one replace the other. When computing aggregable attributes, however, it currently traverses only one of them; extending path search to use all parallel joins is left to future work.

\subsection{Typing: storage, role, and evidence}
\label{sec:types}

\textsc{Tytan} distinguishes a column's storage type from its analytic role. The storage type describes how the values are represented in the database. The analytic role describes how the column should be used by downstream analytic systems. Storage type can usually be determined from metadata and observed values, while analytic role requires some interpretation of what the values mean. 

When the LLM proposes an attribute, it also assigns an initial analytic role. Later passes compare this assignment with the database evidence. Two rules guide this process:
\begin{itemize}
\item \textbf{Evidence may refine, but not override, semantics.} If the model assigns a Numeric role and the sampled values are numeric, \textsc{Tytan} may refine the storage type from string to integer or float. However, a column is not assigned a Numeric role simply because its values contain digits. Zip codes, department codes, and phone numbers are still categorical values and should not be aggregated.
Conversely, the system corrects assignments that clearly conflict with the data. For example, a Numeric role over non-numeric values (course hours recorded as ``10:30--12'') downgrades to a Categorical type. The same applies when a column is labeled Datetime  by the LLM because of its name but contains no temporal values. For example, a column named \code{time\_period} whose values include ``Afternoon Classes'' is treated as Categorical. Declared date and time types are handled differently because explicit database metadata is stronger evidence than a limited sample of values. For example, a date column stored as 20080424 samples as plain integers; the declared DATE type settles what the values alone cannot.

\item \textbf{Computable facts are computed directly from the database.} The declared column type (the database's own metadata) settles the storage type when it is present. An identifier the LLM guessed as a string is recorded as an integer when the database declares it one, and a Numeric attribute declared INTEGER is kept an integer rather than widened to a float. These corrections apply to storage only; the analytic role assigned by
the LLM is never changed here, so an integer-stored zip code stays Categorical and no numeric operation is licensed on it. Because dynamically typed databases can hold values that contradict their declarations, every declaration is checked against the observed values before it is trusted.

Date and time columns are detected from declared metadata when available. If such information is not available, it's identified through value patterns. Their granularity is preserved, so values such as \code{2008-04-24 10:00} are recorded as timestamps rather than dates. Boolean values are normalized, and columns with identifier-like names are checked for common mistakes, such as treating a phone number as a measure. Attribute descriptions may include values sampled directly from the database. The LLM is not allowed to invent these examples.

\end{itemize}

Finally, \textsc{Tytan} verifies each entity's identifier. Entity identifiers are verified separately because it's too consequential: every count, join, and group-by downstream depends
on it. A declared primary key is accepted as the entity key. When the LLM proposes a key for a table without a declared primary key, \textsc{Tytan} checks whether the values are unique. A plausible name is not enough. For example, \code{player\_name} may appear to identify a player while still containing several duplicate names.
When no declared or proposed key can be verified, \textsc{Tytan} searches for other candidates. It considers individual columns from both the entity attributes and the original table, since identifier columns used only in joins may not appear as attributes. Columns with id-like names are checked first, while Numeric measures are excluded. \textsc{Tytan} also considers declared foreign-key columns and, for fact tables, likely composite identifiers such as a combination of dimension keys and a date. Each candidate is tested against the database, and a key is accepted only if it uniquely identifies the rows. Heuristic choices are used only when no candidate can be verified.
This verification ensures that an entity key actually identifies individual rows. It also helps determine whether a table should be represented as an entity at all. If no declared or inferred key uniquely identifies the rows of a junction table, \textsc{Tytan} leaves it as a relationship rather than promoting it to an entity.

\subsection{Naming and surface forms}
After entity and attribute detection, the LLM generates machine-readable names, human-readable labels, and reference templates. \textsc{Tytan} then checks and cleans up these outputs. Abbreviations are expanded. For example, \code{cont\_date} becomes ``Containment Date.'' Additionally, when a proposed label does not match the values in the column, the system derives a new one from the data. For example, a column labeled ``\code{CustomerDetails}'' whose values are customer names is renamed ``\code{CustomerName}.'' Attributes brought in through joins are also given
shorter names, such as ``\code{ClaimFilingDate}'' instead of ``\code{PolicyIDClaimDateClaimMade}.'' Any duplicate names are resolved using deterministic rules.
Reference templates are checked separately. A reference is the short
human-readable label used when a result mentions a specific row, and it
is distinct from the entity identifier. For example, \code{CustomerID} is what
uniquely identifies a customer, but \code{\{CustomerName\}} is how we would reference that entity in a sentence. Each placeholder in an entity's reference template must correspond to one of the entity's attributes. If a template is missing or invalid (for example, the model omits it, or its placeholder names an attribute that was renamed or dropped during assembly) \textsc{Tytan} falls back to the entity identifier, so every entity instance will always be given a valid label.

\subsection{Uncertainty and the clarification loop}
\label{sec:clarification}
Not every decision can be resolved from the database alone. Some depend on the user and the kind of questions they want to ask. An enrollment table, for example, might be treated as its own entity or as a junction between students and courses. If the user wants to ask about enrollments themselves (``How many enrollments were dropped this semester?''), the table should become an entity, with attributes like grade and enrollment date of its own. If the user only wants to reason about students and courses (``Which course has the most students?''), the same table is better treated as a junction, a link between two entities. It supplies the join path, and its columns surface as aggregable attributes on the entities it connects.

\textsc{Tytan} identifies cases like this and asks a focused question when the evidence is not enough. A question might be about a table's role, about a column an attribute could map to, or about a type the values leave unclear. The system might ask, for example: ``Is \code{course\_instructor} a standalone entity, or a junction table connecting courses to instructors?'' The answer goes back through the same pipeline and the same checks. If a table is reclassified, \textsc{Tytan} rebuilds that part of the schema, adds the relevant columns, assigns their types, and updates the joins. Sometimes an answer surfaces a new ambiguity, and the system asks a follow-up. Once everything is resolved, the final ring is assembled.
The user may also skip the questions entirely. Each one has a default answer, so the pipeline can complete without any user input.

\section{Evaluation}
\label{sec:eval}


A \emph{good} schema should be able to model real world entities and relationships, while conveying the semantics to help in querying the dataset. Two schemas will be different in what they model, but will remain the same functionally. We divide the evaluation into 3 key capabilities and test \tytan{} generated schema against the following Research Questions (RQ): 

\begin{itemize}
  \item \textbf{RQ1 (Coverage):} Does the generated schema capture the relevant entities and attributes in the dataset? 
  \item \textbf{RQ2 (Retrieval Correctness):} Are the mappings/sources  to the tables, columns and join paths in the generated schema reachable and correct?
  \item \textbf{RQ3 (Characterization Accuracy):} Are the attributes and entities aligned semantically to the analyses that can be performed on them? For example, can a given attribute really be grouped, averaged or filtered according to the assigned attribute type?
\end{itemize}

These questions define what we \emph{want} to test against, independent of how each is \emph{measured}. Coverage is scored against expectations annotated outside the system: reference schemas written by domain-literate experts and frozen before generation (\S\ref{sec:eval-coverage}, \S\ref{sec:eval-blind}); Retrieval Correctness is scored against the live database by executing the ring instructions (\S\ref{sec:eval-retrieval}); and Characterization Accuracy is scored against the same references (\S\ref{sec:eval-characterization}). A common audit instrument operationalizes all three (\S\ref{sec:eval-intrinsic}).

A note about the scope: we do \emph{not} evaluate the correctness of the downstream analyses. That is covered by the analytics system \satyrn{} \citep{sterbentz2024satyrn}.
\textsc{Tytan}’s contract ends with section \S\ref{sec:ring}. 
We evaluate whether the joins run in the correct direction, whether the column mappings are correct and relevant, the accuracy of the characterization, and the overall semantic meaning of the dataset.

\subsection{Datasets and ground truth}
\label{sec:datasets}

We evaluate on 8 relational databases, summarized in Table~\ref{tab:datasets} (full descriptions and sources in Appendix~\ref{app:data}).

Seven of the eight datasets have an expert-corrected reference schema. For these, we annotated the schema ourselves, learning the data through queries and deciding what the schema should look like. Four are real-world domains: wildfire incidents, school shootings, housing rents, and regional income. The other three come from the Spider text-to-SQL benchmark \citep{yu2018spider}, chosen for their multi-table join structure.

In addition, we chose a live FIFA World Cup 2026 dataset with 10 CSV files and \emph{no} declared primary or foreign keys. This serves as a fully blind generalization test: the expectation suite (\S\ref{sec:eval-coverage}) was annotated and frozen before \tytan{} ever ran on it. Because the source updates daily, we pin a snapshot for our experiments. Results appear in the Blind Generalization section (\S\ref{sec:eval-blind}).

The real-world domain datasets come from the \textsc{Satyrn} evaluation, whose schemas were manually-annotated. The ground-truth rings are the \emph{expert-corrected} versions: script-derived, then reviewed by hand, verifying each mapping and running queries against the data. We treat these rings as a directive to \emph{which} entities, attributes, and relationships matter, and we resolve any structural disagreements we find between the dataset and the generated schema. Some of these surfaced during the audit, for example SSN got misclassified as Numeric instead of being an Identifier type.

\subsection{RQ1: Coverage: Entity and feature}
\label{sec:eval-coverage}
We measure coverage against two independent expectations. First, we check if entities and attributes that are referenced in the ring are reachable or not and then we identify if any relevant entity, feature or relationship was missed. The second is harder, because it depends on semantic judgment suggested by an LLM and is easier to overlook.
\begin{itemize}
    \item \emph{Reference Coverage}: Every entity, attribute, aggregable attribute, and relationship in the ground-truth ring should have a counterpart in the \textsc{Tytan} ring. We match on the underlying column and relationship mappings rather than the names, so a match confirms that both rings point to the same data even when they label it differently.

    \item \emph{Independent expectation suite}: We use this specification on what entities and attributes should be contained in a good schema, annotated from the data description alone (\S\ref{sec:eval-blind}). No table name, column name, or any reference to the schema was given in the description. The suite has 5 independent annotators, two human, domain-literate annotators and a panel of three LLMs (Claude Opus and Sonnet via an agentic CLI in one-shot mode from an empty directory, and gpt-5.5 via the Codex CLI; exact versions recorded in the frozen inputs).
\end{itemize}

The human annotators  were not given access to the dataset or the schema. For the blind FIFA test, one expectation list (the entities, attributes, and relationships a correct schema would contain) was annotated and frozen before generation, the second list is generated after the first run and is recorded with weaker attested blind status, this distinction is preserved in the frozen artifacts.
The LLM annotators were blind by construction, but share a potential bias. Tytan’s own generator is an LLM reading the same description, and one of the panel LLMs belong to the same model family. So LLM annotated lists partly measure the inter-model convergence.
We therefore treat the human list as ground truth. Then we report the per-annotator recall alongside the panel consensus  ($\geq$2 of 3) \citep{panickssery2024selfpreference}.

The results are scored as the following:
\begin{itemize}
    \item \emph{satisfied} - ring matches semantically with the human adjusted spot checks 
    \item \emph{missed} - present in the database but missed in the ring 
    \item \emph{unsatisfiable} - not in the data at all; excluded from the recall denominator and are reported separately since it is measured against the dataset rather than the system.
\end{itemize}
Across all seven reference domains, \textsc{Tytan} reaches every entity, attribute, and aggregable attribute in the reference ring, with no misses (Table~\ref{tab:coverage}). The eighth database (FIFA WC2026) has no expert reference ring. Its coverage is measured against the independent blind expectation suite and reported separately (\S\ref{sec:eval-blind}).

\begin{table}[t]
\centering
\small
\begin{tabular}{lcccc}
\toprule
\textbf{Domain} & \textbf{Ent.} & \textbf{Attrs} & \textbf{Aggr.} \\
\midrule
college\_3 & 4/4 & 40/40 & 60/60 \\
insurance\_policies & 5/5 & 42/42 & 24/24 \\
hospital & 10/10 & 60/60 & 169/169 \\
income\_disparity & 1/1 & 14/14 & --- \\
zillow & 2/2 & 7/7 & 2/2 \\
school\_shooting & 4/4 & 126/126 & 19/19 \\
wildfire & 1/1 & 16/16 & --- \\
\bottomrule
\end{tabular}
\caption{RQ1 reference coverage: every reference entity, attribute, and
aggregable is reachable in the generated ring (provenance-matched,
join-equivalence closure).}
\label{tab:coverage}
\end{table}

\subsection{RQ2: Self-generated retrieval tests}
\label{sec:eval-retrieval}
A ring is operationally a set of instructions, defining the mappings of the entities to tables, column to attributes, and join paths to relationships. 

The Retrieval Correctness tests these instructions by executing them. For each ring we \emph{self-generate} a test suite which includes: 
\begin{itemize}
    \item For each entity, sample instances by the declared ID and confirm the id is non-null and unique over all rows.
    \item For each attribute, retrieve its value using the suggested mapping in the ring and test the declared table and columns exist and are reachable.
    \item For each aggregable attribute, traverse through the suggested join path and verify if the retrieval is successful and the resulting rows correspond to the underlying relationships
    \item For every reference template, render the labels for the sampled instances, and verify  if they are semantically correct.
\end{itemize}
The tests are structured to scale with the ring and are independent of how detailed the annotations for a database are. Every claim made in the ring is tested directly with the database. The  (Table~\ref{tab:retrieval}) shows the results for each database against each claim made in the ring and the percentage of how many were actually found in the database. 

The test suite included four test classes:
\begin{itemize}
\item \emph{Identifier uniqueness:} non null and distinct over all rows,

\item \emph{Attribute mapping:} declared tables and columns exist and are reachable,

\item \emph{Join executability:} every attribute join path connects correctly and executes, and

\item \emph{Reference resolution:} declared labels resolve to an attribute, including the engine synthesized IDs.
\end{itemize}

\begin{table}[t]
\centering
\small
\begin{tabular}{lcc}
\toprule
\textbf{Domain} & \textbf{Claims} & \textbf{Pass} \\
\midrule
college\_3 & 325 & 100\% \\
insurance\_policies & 145 & 100\% \\
hospital & 854 & 100\% \\
income\_disparity & 16 & 100\% \\
zillow & 19 & 100\% \\
school\_shooting & 290 & 100\% \\
wildfire & 38 & 100\% \\
\midrule
FIFA WC2026 (blind) & 2{,}071 & 100\% \\
\bottomrule
\end{tabular}
\caption{RQ2 self-generated retrieval tests: every claim the ring makes
(ID uniqueness, attribute mapping, join-chain execution, reference
resolution) executed against the live database.}
\label{tab:retrieval}
\end{table}

The suite also catches key-selection errors that coverage misses. In one case, \textsc{Tytan} defaulted to an arbitrary column when no real key was found. In another, it keyed an entity on a name column with duplicate values, because its computational time budget ran out on measure columns before it reached a real key candidate. The identifier uniqueness test (\S\ref{sec:types}) flagged both against the live data, and each is handled by general, deterministic logic (an identifiability guard, and better ordering of key candidates) rather than data-specific patches.

Because these tests are generated from \textsc{Tytan}'s own output, they cannot catch omissions, that's why we do a coverage evaluation on entities. Retrieval Correctness checks that everything the ring \emph{claims} is executable and truthful.

\subsection{RQ3: Characterization accuracy}
\label{sec:eval-characterization}

Downstream analysis decisions rely on semantic roles. For example, a consumer can average a Numeric attribute, group by a Categorical attribute and build a time series on a Datetime attribute. We measure the accuracy of the categorization of these attributes against the ground truth rings, matching the mappings for the attributes using two criteria: role agreement (on types of Attribute, Numeric, Categorical, Numeric, Datetime) and storage-type agreement (declared column types, value samples) with disagreements against the instance evidence. Table~\ref{tab:characterization} reports the results for the Characterization Accuracy test. It found two misclassified attributes in the first round, (declared integer in the database but were datetime) but were semantically corrected after the checks and verified at 100\%.

\begin{table}[t]
\centering
\small
\begin{tabular}{lcc}
\toprule
\textbf{Domain} & \textbf{Roles} & \textbf{isa (family)} \\
\midrule
college\_3 & 93.9\% & 98.0\% \\
insurance\_policies & 100\% & 100\% \\
hospital & 94.7\% & 100\% \\
income\_disparity & 100\% & 100\% \\
zillow & 100\% & 100\% \\
school\_shooting & 100\% & 100\% \\
wildfire & 100\% & 100\% \\
\bottomrule
\end{tabular}
\caption{RQ3 Characterization accuracy on the mapped attributes in the ring vs the database.
There were 21 miscategorized  attributes found, where identifier type numerics were found as summable metrics instead of categorical}
\label{tab:characterization}
\end{table}

\subsection{Blind generalization test}
\label{sec:eval-blind}
We used the FIFA World Cup 2026 dataset for held-out blind test: a live dataset with ten tables, downloaded as CSVs with no declared primary or foreign keys. The expectation suite (\S\ref{sec:eval-coverage}) was annotated and committed before \textsc{Tytan} saw it. 

The first run was executed fully autonomously with no clarification answers. In this run, the entities (\code{Teams}, \code{Players}, \code{Matches}, \code{Venues}, \code{Referees}, and \code{Tournament Stages}) matched the core of all the four frozen expectation lists. \textsc{Tytan} was able to infer the correct identifier for all the entities despite the absence of any declared keys in the dataset.

But it also failed in two instructive ways: three spurious joins passed LLM verification (goal/save counts whose small-integer ranges overlap sequential ids), and it also missed three foreign keys. These false joins are the more telling failure: the LLM approved them itself, which is why join validation cannot rest on the model alone. The keys went undiscovered as the entities created by the iterative sweep were invisible as reference targets and a second foreign key between an already connected table was dropped. Since these failures were held-out rather than cherry-picked, their diagnosis is the evidence for the method. Each resolution is deterministic and dataset independent, and can be regression-tested. We then ran the fixed pipeline again, still with no clarification answers. It gave us 7 entities, with verified keys, 19 joins and a 100\% retrieval test pass rate over 2,071 self-generated claims 
(Table~\ref{tab:retrieval}).

The frozen expectation by the second annotator aligns with a failure from the first run. In this data, a match in the \code{matches} table involves two teams, a home side and an away side, and has two columns referencing the same \code{Teams} table as foreign keys. The annotator’s frozen expectation list for the \code{match} entity, independently expected TeamA and TeamB as attributes, i.e. it demanded the two-references-to-one-table structure that \tytan{} initially missed. The annotator arriving at this requirement, based only on the dataset description and no schema shows that handling this structure correctly is genuinely a part of what a correct schema needs.

The final rings, scored against all the frozen expectation lists (Table~\ref{tab:expectations}) yield 100\% recall over all the satisfiable items that the annotators had in their expectation list. The unsatisfiable items were those that did not exist in the dataset and there were 41 of these. We do not consider this as the system failure but gaps in the data itself. We designed the evaluation to separate these concerns. Similarly, all three LLMs in the annotator panel, expected a “Tournament” entity which did not exist in the dataset. Judgments were produced by an LLM judge \citep{zheng2023judging} (pinned \code{gpt-5.5}), where each citation was validated against the ring and the dataset and then is rechecked by a human. The rechecked items, 108 in total, included,  every unsatisfiable item, judgment calls, both cases where the judges disagreed with consensus and a 20\% sample of satisfied verdicts.

The human-adjudiction overturned 13 of the LLM judges' verdicts, and every one was in the same direction: the judge had excluded the item as unsatisfiable for the data did not exist. But on re-check it was found that it was deliverable based on one consistent rule: the data is in the database, and the question can be answered through the schema then the item is delivered, counting and aggregation come under the analytics’s engine scope.

\emph{The paradigm case:} The annotators expected “tournament year” and “start/ end dates”, but no such exact columns exist. But the schema delivered \code{Match.Date} as datetime and year filtering, calendar rollups, start/end dates are derivations the user computes. The schema’s obligation is to deliver the correct date type (\S\ref{sec:ring}). The judge thus marked these as unsatisfiable and the human corrected them.  The rings scored here were generated fully autonomously and without any clarification answers. Table~\ref{tab:expectations} reports recall after this human review of the judge's scores.

\begin{table}[t]
\centering
\small
\setlength{\tabcolsep}{4pt}
\begin{tabular}{lcccc}
\toprule
\textbf{Annotator} & \textbf{Ent.} & \textbf{Feat.} & \textbf{Rel.} & \textbf{Unsat.} \\
\midrule
Claude Opus & 6/6 & 41/41 & 8/8 & 4 \\
Codex (\code{gpt-5.5}) & 8/8 & 59/59 & 12/12 & 16 \\
Claude Sonnet & 10/10 & 49/49 & 10/10 & 14 \\
Human \#1 (pre-gen.) & 4/4 & 14/14 & 2/2 & 4 \\
Human \#2 (att.-blind) & 3/3 & 12/12 & 3/3 & 3 \\
\midrule
Panel consensus & 7/7 & 44/44 & 10/10 & 7 \\
\bottomrule
\end{tabular}
\caption{Blind expectation-suite recall on FIFA WC2026, after human
adjudication: satisfied / satisfiable items per class (entities,
features, relationships), plus unsatisfiable items (absent from the data;
excluded from the denominator). Recall is 100\% for every annotator and the
$\geq$2-of-3 panel consensus; no expectation item present in the database 
was missed.}
\label{tab:expectations}
\end{table}

\subsection{Cross-run stability}
\label{sec:eval-stability}

Nondeterminism in LLMs is a legitimate concern. A tool that produces a different schema every time it is run would be hard to trust. \tytan{}, however, is insulated from this by dint of its architecture: 1) the fixed vocabulary of the semantic layer (e.g., entity, attribute, join, etc.) constrain the LLM generation; 2) the deliberate inclusion of the human in the loop ensures the schema remains recognizably consistent with the data.

We nonetheless evaluate nondeterminism in \tytan{} with three fully-autonomous runs on the 4 multi-table domain datasets (hospital, college, school shootings, FIFA) on identical inputs and code. We compare on these on the structural decisions that matter, joins and entities that were identified, which keys were inferred and how columns were characterized. 

We claim that the grounding architecture makes these decisions deterministic and is checked against the actual dataset. Our results show that the join sets agree on every pair of run in every domain (Jaccard 1.0), including the FIFA dataset where every join needs to be inferred. The entity IDs are identical across all runs (25/25 entity-table pairs), as are entity sets, with a single exception where one run treated grade-conversion chart as an entity while others treated it as a lookup. Crucially, this is “information-equivalent” -- either way, the same data is accessible through the schema. It is a labeling choice not a functional difference. 

Shape and identifiability guards make these structural decisions deterministic across fully autonomous runs. 
Residual variance is now confined to semantic typing of the few columns found in the datasets (pairwise role agreement 94–100\%), e.g. a building-floor-number, ages, elevation-column, cases where the judgment calls vary as the data alone does not settle the answer.

\subsection{Metadata ablation}
\label{sec:eval-ablation}
The blind test’s clean results on  FIFA WC-2026, with zero metadata (no declared primary keys, foreign keys, or column types) raises a natural question: how much do declared constraints actually contribute? We test this directly on the Spider datasets (insurance, college, hospital), building its degraded twin, by stripping away their metadata. We produced a CSV-equivalent copy with identical tables and rows and values and then ran \tytan{} on both versions (original and the metadata-stripped), with identical code, fully autonomously, making sure metadata is the only variable. The original database’s declared foreign keys serve as the ground truth for joins, but \tytan{} now need to rediscover them from the values alone for the metadata-stripped dataset (Table~\ref{tab:ablation}).
Results:
\begin{itemize}
    \item Core Structure survives metadata loss: 

The \emph{structure is metadata-robust} and stays intact: entity sets, attribute coverage, verified keys (identical on 18 of 19 shared entity tables) and analytic roles remain unchanged. The insurance database’s stripped run reproduces the metadata run’s join graph exactly.
    \item Joins are where the metadata matters:
    
\emph{Joins are where metadata has benefits, and the driver is how much the identifier value ranges overlap}. 
Value-only discovery works near-perfectly where identifiers are distinct: insurance recovered 4/4 foreign keys, college 8/9. On the otherhand, in the hospital dataset, a dozen overlapping small integer code spaces drops the recovery to 13/25 and four coincidental-overlap joins leak past the guards, which are tuned to catch ranges overlapping three or more key spaces.
So  we see metadata isn't needed for structure but is valuable for joins when the identifier values are ambiguous. 
 \item Failure caught by self-generated tests:
 
With no declared foreign key to suggest composite-key candidates, one entity resorted to an unverified fallback id, which was flagged by the Retrieval Correctness tests. \emph{Storage-hint agreement} drops by design rather than by error, as all the columns in the metadata-stripped datasets are declared as TEXT, but the semantic roles are held at 92-100\% throughout.

\end{itemize}
\begin{table}[t]
\centering
\small
\begin{tabular}{llccc}
\toprule
\textbf{Domain} & \textbf{Arm} & \textbf{FK rec.} & \textbf{Spur.} & \textbf{RQ2} \\
\midrule
insurance (4) & meta & 4/4 & 0 & 145/145 \\
 & stripped & 4/4 & 0 & 138/138 \\
college (9) & meta & 9/9 & 0 & 331/331 \\
 & stripped & 8/9 & 0 & 403/403 \\
hospital (25) & meta & 25/25 & 0 & 854/854 \\
 & stripped & 13/25 & 4 & 589/590 \\
\bottomrule
\end{tabular}
\caption{Metadata ablation: FK recovery(against the database’s constraints), spurious joins and self generated tests with meta data stripped (“stripped”) vs intact (“meta”). Entity coverage and attributes are covered at 100\%, ids and analytic roles are mostly unchanged for both “meta” and “stripped”(92\%-100\%); The college counts exclude the two genuinely undeclared references both arms discover. The single RQ2 failure is the composite-key fallback discussed in the text.}
\label{tab:ablation}
\end{table}

\subsection{Supporting instrument: artifact-grounded audit}
\label{sec:eval-intrinsic}
Alongside the scored tests above, we ran a second defect audit as an extra check. Each ring is checked against the same deterministic \emph{ground-truth artifact} used throughout: tables, columns, declared keys, value samples, cardinalities, and null rates, all pulled directly from the database. The artifact uses no LLM judgment, and every issue the audit flags must point to evidence in it, not to the live database or the model's memory. To limit single-model bias, three models audit each ring on their own and a fourth checks the merged report. Findings follow a fixed defect taxonomy (Table~\ref{tab:taxonomy}) whose codes map onto the three research questions: D2/D5 to coverage, D1/D3/D6/D7 to retrieval, and D4/D9 to characterization, with D0/D8 non-functional. The full rubric, severity levels, and anti-pattern examples are in Appendix~\ref{app:rubric}.

\begin{table}[h]
\centering
\small
\begin{tabular}{llc}
\toprule
\textbf{Code} & \textbf{Defect class} & \textbf{RQ} \\
\midrule
D0 & Structural validity & --- \\
D1 & Join fidelity (missing/spurious) & RQ2 \\
D2 & Entity coverage & RQ1 \\
D3 & Primary-key correctness & RQ2 \\
D4 & Attribute type correctness & RQ3 \\
D5 & Attribute coverage & RQ1 \\
D6 & Reference-template validity & RQ2 \\
D7 & Aggregable correctness & RQ2 \\
D8 & Redundancy (non-functional) & --- \\
D9 & Internal consistency & RQ3 \\
\bottomrule
\end{tabular}
\caption{Audit defect taxonomy and its mapping onto the functional
research questions. Each finding carries a severity (critical / major /
minor) and must cite artifact evidence.}
\label{tab:taxonomy}
\end{table}

\subsection{Downstream Demonstration}
\label{sec:eval-extrinsic}

\textsc{Satyrn} is a platform for analytics-augmented generation that operates on the entities and attributes specified in the rings \textsc{Tytan} produces. Analytics within \textsc{Satyrn} are defined at the ring level and operate on specific entities and their attributes. The analytics engine of \textsc{Satyrn} will then execute these analytics and produce the necessary SQL queries against the underlying database. These method of carrying out analytics has been used to create a natural language notebook in which users can search for and select questions to answer in a turn-by-turn manner \cite{paley2021data}, generate high-quality training data for improving the text-to-SQL capabilities of large language models via reinforcement learning \cite{sterbentz2026ringsql}, and answer questions with reports that provide contextualizing information for a user's question \citep{sterbentz2024satyrn}.

\begin{figure}[t]
\centering
\resizebox{0.98\columnwidth}{!}{\begin{tikzpicture}[
    every node/.style={font=\sffamily\small},
    leftbubble/.style={
        rectangle, rounded corners=8pt, fill=bubbleLeft,
        text=textLeft, align=left, inner sep=8pt, anchor=north west,
        drop shadow={opacity=0.15, shadow xshift=0.5pt, shadow yshift=-0.5pt}
    },
    rightbubble/.style={
        rectangle, rounded corners=8pt, fill=bubbleRight,
        text=textRight, align=left, inner sep=8pt, anchor=north east,
        drop shadow={opacity=0.15, shadow xshift=0.5pt, shadow yshift=-0.5pt}
    }
]

\draw[rounded corners=18pt, fill=white, draw=gray!40, line width=1pt]
    (-0.6,-9.5) rectangle (7.9,0.0);

\node[rightbubble, text width=4.0cm, yshift=-\gap] (m2) at (\rightx,0.0) {What are the top five cities in Illinois by school shootings?};
\path let \p1=(m2.south) in coordinate (row1) at (0,\y1);

\node[leftbubble, text width=4.2cm, yshift=-\gap] (m3) at (\leftx,0|-row1) {The top five cities in Illinois by school shootings are as follows:

1. Chicago - 63 incidents

2. Rockford - 6 incidents

3. Champaign - 4 incidents

4. Springfield - 3 incidents

5. Peoria - 3 incidents

Chicago has significantly more incidents than the other cities listed.};
\path let \p1=(m3.south) in coordinate (row2) at (0,\y1);

\node[rightbubble, text width=3.6cm, yshift=-\gap] (m4) at (\rightx,0|-row2) {How many fires were there in Illinois in 2020?};
\path let \p1=(m4.south) in coordinate (row3) at (0,\y1);

\node[leftbubble, text width=4.2cm, yshift=-\gap] (m5) at (\leftx,0|-row3) {In 2020, Illinois experienced a total of 125 wildfires.};

\end{tikzpicture}}
\caption{Example end-to-end system interaction}
\label{fig:qa_example}
\end{figure}
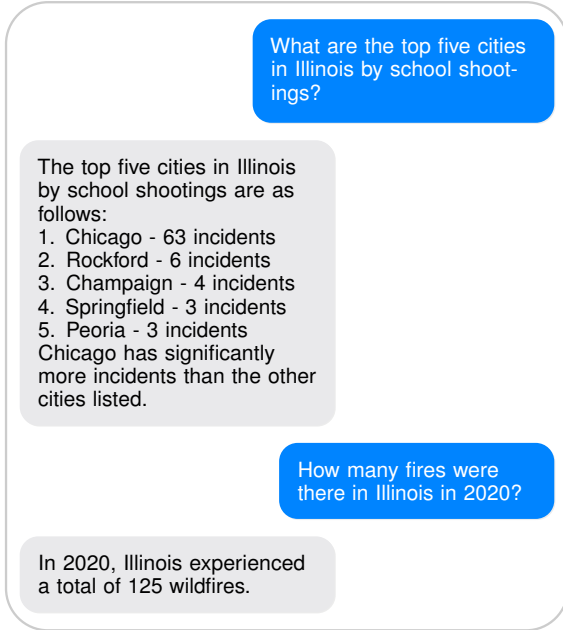

For this last application, we demonstrate the generated schema end-to-end:  \textsc{Satyrn}, loaded with the \textsc{Tytan} rings school shooting and wildfire, answers analytic questions over the live databases (Figure \ref{fig:qa_example}).
We emphasize that this demonstration validates integration and not analysis quality. The analytic correctness is the consuming system’s evaluation not the schema’s.

\section{Discussion}
\label{sec:discussion}
\paragraph{Deterministic grounding.} Most errors we encountered during development occurred when an LLM output was accepted without enough support from the database. Some example failures include: the model proposing joins because two columns had similar names, including example values that were not present in the data, inferring types from column names alone, and selecting keys without checking whether they were unique. Better and more strict prompts did not reliably prevent these errors. Instead, we moved verifiable decisions out of the LLM: values, uniqueness, and column overlap are computed directly, while the model is used for semantic judgments that require domain knowledge. Those judgments are then checked against the database before they are added to the schema. This separation between computed facts and model-based interpretation, together with the defect taxonomy used to audit the output, may also be useful in other LLM-based data-engineering systems.

\paragraph{Results.} The results support two conclusions. First, removing the declared metadata had little effect on the entities, attributes, keys, and roles produced by the system. This suggests that the approach can work with raw CSV exports as well as curated relational databases. Second, Join discovery was more sensitive to missing metadata. When several identifier columns contained heavily overlapping value ranges, the data alone could not always distinguish a real relationship from a coincidental match. In these cases, \textsc{Tytan} will refer to user input or declared constraints to settle them.

\section{Conclusion}
In this paper we presented \textsc{Tytan}, an interactive neurosymbolic system designed for constructing analytic semantic schemas from relational databases. The system combines LLM-based interpretation with deterministic checks against the database instance and asks the user to resolve cases that cannot be settled from the data alone. Its output is a self-contained semantic schema that can be used by downstream analytic tools. We also introduced an audit methodology for evaluating generated schemas and tested whether the resulting schemas support downstream analysis. Across eight databases, including a blind test with no declared keys, the generated schemas covered every feature in the expert-corrected references and successfully executed all 3{,}758 self-generated retrieval tests against the source data. They also satisfied all satisfiable expectations written by five independent blind evaluators. These results suggest that checking model-generated decisions against the database can make automated semantic schema construction substantially more reliable.

\section*{Limitations}
\textsc{Tytan} was developed for relational and tabular data. It can process hierarchical formats such as JSON or event logs only after they have been flattened, which may discard some of their original structure. Graph-shaped sources such as knowledge graphs are outside the current scope of the system. They already represent entities and relationships explicitly rather than through tables, columns, and joins, and therefore require a different approach to schema discovery \citep{kelloumenouer2022survey}. Join discovery is intentionally conservative, which means \textsc{Tytan} may miss real relationships when the two sides use different representations. For example, one table may store a numeric identifier while another stores the corresponding name. Additionally, the value-overlap checks assume that \textsc{Tytan} can inspect the full database. When only a sample is available, or access to the data is limited, finding no shared values does not rule out a relationship confidently. In such cases, checks that rely on overlap become less reliable. The metadata ablation evaluation in (\S\ref{sec:eval-ablation}) shows where join recovery begins to break down. It remains strong when identifier spaces are easy to distinguish, but becomes less reliable when many columns use overlapping small-integer values. Removing metadata also increases construction cost by roughly a factor of five. Large schemas also increase the number of candidate joins that require LLM verification and may therefore raise both runtime and cost. Composite-key discovery uses declared foreign keys to generate candidate combinations, so it is less effective when those declarations are missing. 

The clarification loop assumes that the user understands the domain, although no database expertise is required. When no user is available, the system applies default AI-suggested answers. We evaluate the overall performance of this unattended setting but do not measure the quality of each default separately. Because a human is part of the pipeline, the clarification component deserves its own evaluation, which we acknowledge is part of future work.
Finally, the expert reference schemas were written for the same downstream system used in our extrinsic evaluation, \textsc{Satyrn}. Although the ring contains the kinds of information commonly required by semantic-layer systems, its usefulness for other consumers has not yet been tested directly.

\section*{Acknowledgments}
We would like to thank the Center for Advancing the Safety of Machine Intelligence (CASMI) for funding this work.

\bibliography{tytan}

\appendix

\section{Datasets}
\label{app:data}
Table~\ref{tab:datasets} summarizes the eight datasets used in our evaluation. Four represent real-world analytic domains for which the original rings were written by hand. We also include three databases from the Spider text-to-SQL benchmark \cite{yu2018spider}. They were selected because they contain non-trivial relationships across multiple tables. The final database was reserved for the blind evaluation.

\textbf{Wildfire Occurrence} The Wildfire Dataset is an extensive collection of data which contains records for 2.3 million georeferenced wildfires in the United States from 1992 through 2020, representing approximately 180 million acres burned \citep{short2022wildfire}. We host the database in PostgreSQL.

\textbf{School Shooting Incidents} The K-12 School Shooting Database,\footnote{\url{https://k12ssdb.org}}, originally developed at The Center for Homeland Defense and Security, covers incidents at U.S.\ schools from 1970 through June 2022. It includes any incident in which a gun was brandished or fired, or a bullet struck school property, whether or not anyone was injured or killed.

\textbf{Zillow Observed Rent Index} The Zillow Observed Rent
Index\footnote{\url{https://www.zillow.com/research/data/}} is a rental price index designed to accurately represent the entire rental housing market, rather than just the properties currently listed for rent. This index averages the rents rents in the 40th--60th percentile of each region's housing stock \citep{zillow-group-inc-2023}.

\textbf{Income Disparity} This database contains information about the earnings of residents of specific geographic locations, such as counties or metropolitan areas \citep{us-bureau-of-economic-analysis-personal-income-2022,us-bureau-of-labor-statistics-unemployed-2023, us-census-bureau-age-17-2022, us-census-bureau-median-household-2022, us-census-bureau-all-age-2022}.

\textbf{Spider databases} We use three databases from the Spider text-to-SQL benchmark: college, hospital, and insurance \citep{yu2018spider}. 

\textbf{FIFA World Cup 2026 (blind)} The blind-test database is a live, relationally normalized FIFA World Cup 2026 dataset with ten tables and no declared primary or foreign keys of any kind. The source updates daily.

\begin{table*}[ht]
\small
\centering
\begin{tabular}{p{0.16\linewidth}p{0.19\linewidth}p{0.55\linewidth}}
\toprule
\textbf{Domain} & \textbf{Dataset} & \textbf{Description} \\
\midrule
Environmental sustainability & Wildfire Occurrence & 2.3M geo-referenced
U.S.\ wildfire records, 1992--2020 \citep{short2022wildfire}. \\
Public safety & School Shooting Incidents & K-12 school shooting
incidents, 1970--June 2022. \\
Urban housing & Zillow Rent Index & Regional rent-index time series;
the fact table declares no key. \\
Socioeconomic & Income Disparity & BEA personal income by county and
metro area, with BLS and Census series. \\
Education & College (Spider) & Students, courses, an enrollment
junction, and a grade-conversion lookup; 9 declared FKs
\citep{yu2018spider}. \\
Healthcare & Hospital (Spider) & 25 declared FKs; overlapping
small-integer code ranges (the ablation stress case). \\
Insurance & Insurance policies (Spider) & Five entity tables; 4
declared FKs. \\
Sports (blind) & FIFA World Cup 2026 & Live ten-table dataset with zero
declared keys. \\
\bottomrule
\end{tabular}
\caption{The eight evaluation databases: seven with expert-corrected reference schemas (four real-world domains and three Spider databases) and one held-out blind test.}
\label{tab:datasets}
\end{table*}

\section{Ring Format Reference}
\label{app:ring}

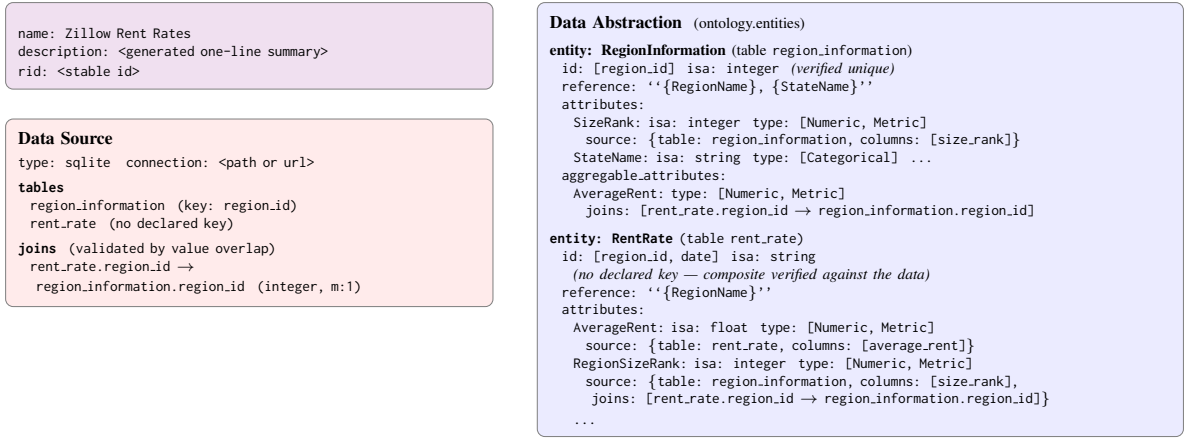
\begin{figure*}[t]
\centering
\resizebox{0.98\textwidth}{!}{
\begin{tikzpicture}[
  font=\scriptsize,
  panel/.style={draw=black!50, rounded corners=3pt, inner sep=5pt,
                align=left},
  head/.style={font=\scriptsize\bfseries}
]

\node[panel, fill=violet!12, text width=64mm] (meta) {%
{\head}\\[2pt]
\ttfamily\tiny name: Zillow Rent Rates\\
\ttfamily\tiny description: <generated one-line summary>\\
\ttfamily\tiny rid: <stable id>};

\node[panel, fill=red!8, text width=64mm, below=4mm of meta] (ds) {%
{\head Data Source}\\[2pt]
\ttfamily\tiny type: sqlite\quad connection: <path or url>\\[3pt]
{\bfseries\tiny tables}\\
\ttfamily\tiny\ \ region\_information\ \ (key: region\_id)\\
\ttfamily\tiny\ \ rent\_rate\ \ (no declared key)\\[3pt]
{\bfseries\tiny joins}\ \ {\tiny\itshape (validated by value overlap)}\\
\ttfamily\tiny\ \ rent\_rate.region\_id \(\rightarrow\)\\
\ttfamily\tiny\ \ \hphantom{\ }region\_information.region\_id\ \ (integer, m:1)};

\node[panel, fill=blue!8, text width=86mm, anchor=north west]
      (ont) at ([xshift=6mm]meta.north east) {%
{\head Data Abstraction\ \ {\mdseries\tiny(ontology.entities)}}\\[3pt]
{\bfseries\tiny entity: RegionInformation}\ {\tiny(table \texttt{region\_information})}\\
\ttfamily\tiny\ \ id: [region\_id]\ \ isa: integer\ \ \rmfamily\tiny\itshape(verified unique)\\
\ttfamily\tiny\ \ reference: ``\{RegionName\}, \{StateName\}''\\
\ttfamily\tiny\ \ attributes:\\
\ttfamily\tiny\ \ \ \ SizeRank:\ isa: integer\ \ type: [Numeric, Metric]\\
\ttfamily\tiny\ \ \ \ \ \ source: \{table: region\_information, columns: [size\_rank]\}\\
\ttfamily\tiny\ \ \ \ StateName:\ isa: string\ \ type: [Categorical]\ \ ...\\
\ttfamily\tiny\ \ aggregable\_attributes:\\
\ttfamily\tiny\ \ \ \ AverageRent:\ type: [Numeric, Metric]\\
\ttfamily\tiny\ \ \ \ \ \ joins: [rent\_rate.region\_id \(\rightarrow\) region\_information.region\_id]\\[4pt]
{\bfseries\tiny entity: RentRate}\ {\tiny(table \texttt{rent\_rate})}\\
\ttfamily\tiny\ \ id: [region\_id, date]\ \ isa: string\\
\ttfamily\tiny\ \ \ \ \rmfamily\tiny\itshape(no declared key --- composite verified against the data)\\
\ttfamily\tiny\ \ reference: ``\{RegionName\}''\\
\ttfamily\tiny\ \ attributes:\\
\ttfamily\tiny\ \ \ \ AverageRent:\ isa: float\ \ type: [Numeric, Metric]\\
\ttfamily\tiny\ \ \ \ \ \ source: \{table: rent\_rate, columns: [average\_rent]\}\\
\ttfamily\tiny\ \ \ \ RegionSizeRank:\ isa: integer\ \ type: [Numeric, Metric]\\
\ttfamily\tiny\ \ \ \ \ \ source: \{table: region\_information, columns: [size\_rank],\\
\ttfamily\tiny\ \ \ \ \ \ \ joins: [rent\_rate.region\_id \(\rightarrow\) region\_information.region\_id]\}\\
\ttfamily\tiny\ \ \ \ ...};
\end{tikzpicture}}
\caption{A snippet ring generated by \textsc{Tytan} for the housing-rent domain, shown in the three-section presentation style used for the expert-annotated rings of \citet{sterbentz2024satyrn}: ring metadata (purple), data source and validated joins (pink), and the data
abstraction layer (blue). Every element shown (keys, joins, types, labels) was generated and verified automatically. RentRate's composite key was verified against the data to be unique against since the table did not have any declared keys.}
\label{fig:ringfull}
\end{figure*}

Figure~\ref{fig:ringfull} shows a shortened ring for the housing-rent domain: two entities built over two tables, connected by one validated join. A ring is a single JSON document with three sections.

The \emph{metadata} section (purple) carries the ring's id (\code{rid}), its name, and a generated one-line description of the ring.

The \emph{data source} section (pink) records how to reach the data:
the connection type, the tables, and the global \code{joins} list. Each join names its two tables, its two columns, and the column type. Every join in this list was validated against the instance
(\S\ref{sec:joins}). There is no separate relationships section:
cross-entity structure lives in this joins list and in the per-attribute join paths below.

The \emph{data abstraction} section (blue) lists the entities found in the dataset. Each entity includes a verified \code{id}, which may be a composite of more than one column, along with a singular \code{name}, and a \code{reference} template of at most two \code{\{Attribute\}} placeholders. Each entity's \code{attributes} include columns from the entity's own table and fields reached through forward many-to-one joins. It's \code{aggregable\_attributes} include fields reachable through reverse or multi-hop relationships, together with the full join path needed to query them. 
Each attribute records a human-readable \code{name}, its physical storage type in \code{isa} (string, integer, float, date, datetime, timestamp, or boolean), and its analytic role in \code{type} (Identifier, Categorical, Numeric, Metric, or Datetime). It also includes a short description of the attribute along with values sampled from the database and the source column from which it was derived.
The Zillow Rent Rate's ring (shown in Figure~\ref{fig:ringfull}) would be stored in a JSON file as: 

\begin{small}
\begin{verbatim}
{"name": "RentRate",
 "id": {"isa": "string", "source":
   {"table": "rent_rate",
    "columns": ["region_id","date"]}},
 "reference": "{RegionName}",
 "attributes": {
  "AverageRent": {"isa": "float",
    "type": ["Numeric","Metric"],
    "source": {"table": "rent_rate",
      "columns": ["average_rent"],
      "joins": []}},
  "RegionSizeRank": {"isa": "integer",
    "type": ["Numeric","Metric"],
    "source":
     {"table": "region_information",
      "columns": ["size_rank"],
      "joins": ["rent_rate.region_id->
        region_information.region_id"]}},
  ...}}
\end{verbatim}
\end{small}

The original ring format required manual annotating \citep{sterbentz2023lightweight}. An expert specified the entities, attributes, and joins in a pre-annotated CSV file, which was then converted into the JSON format used by Satyrn \citep{sterbentz2024satyrn}. Interactive ring construction was left for future work. By contrast, the ring shown in Figure~\ref{fig:ringfull} was generated from a database connection string and a one-sentence description. Before producing the final ring, \textsc{Tytan} checked the proposed keys, joins, and analytic roles against the database.

\section{Audit Rubric}
\label{app:rubric}
The audit rubric checks each ring only against the deterministic database artifact. The auditor does not access the live database or add values that are not contained in the artifact.
Defect classes:

\vspace{0.5em}
\begin{center}
\small
\begin{tabular}{@{}l@{\ \ }p{0.78\columnwidth}@{}}
\toprule
D0 & Structural sanity (required fields; stop on failure) \\
D1 & Join fidelity (declared FKs present; no spurious joins) \\
D2 & Entity coverage (every substantive table reachable) \\
D3 & Primary-key correctness (ids unique and non-null) \\
D4 & Attribute type correctness (isa vs.\ declared type and samples) \\
D5 & Attribute coverage (no substantive column dropped) \\
D6 & Reference-template validity (placeholders resolve) \\
D7 & Aggregable correctness (join paths well-formed) \\
D8 & Redundancy (cosmetic duplicates) \\
D9 & Internal consistency (cross-field contradictions) \\
\bottomrule
\end{tabular}
\end{center}

\vspace{0.5em}

Each defect is then categorized into one of the following three severity classes: \emph{critical} (breaks retrieval or misleads analysis), \emph{minor} (degrades quality), \emph{cosmetic}.

Two rules were added while applying the rubric following typing rules mentioned in \S\ref{sec:types}. The first specifies cases that should not be treated as errors. These include categorical columns stored as integers, date and time granularity supported by the sampled values, and undeclared joins whose value overlap has already been verified. The second rule handles joins that appear plausible from their column names. In an early audit, the reviewer accepted one such join even though the two columns had no values in common. The auditor must also confirm value overlap in the ground-truth artifact of the database before accepting an undeclared relationship.

\section{Clarification Question Types}
\label{app:clarification}
Each clarification question is stored as a structured object that includes a deterministic identifier. This allows for the same inputs produce the same question and previous answers can be reused safely. This object also contains the question text, the model's suggested answer and confidence score, and the context needed to apply the user's response without relying on session state. There are four categories of clarifying questions: 

\begin{itemize}
\item \textbf{table\_mapping}: asks which table corresponds to a proposed entity. The answer updates the entity's source table and rebuilds its attribute mappings.

\item \textbf{column\_mapping}: asks which column(s) correspond to a proposed attribute. The answer rewrites the attribute's provenance.

\item \textbf{satyrn\_type}: asks how an ambiguous column should be used in analysis. For example, ``Should \code{Age} be treated as a numeric measure, allowing average age across students of a class, or as a category for grouping students by age?'' The answer changes the column's analytic role but leaves its storage type unchanged.

\item \textbf{table\_role}: asks how a structurally ambiguous table should be represented. Whether it should be an entity, a junction table, a passthrough lookup, or a technical table to ignore. The user's answer replaces the initial classification and is preserved throughout the remaining stages of the pipeline.

\end{itemize}

\section{Prompts}
\label{app:prompts}
All LLM calls in \textsc{Tytan} use schema-contrained outputs. The released code includes the full system prompts used in our experiments. Their main instructions are summarized below:
\begin{itemize}
\item \textbf{Entity proposal}: \textit{given the table and column names, sampled values, and the user's one-sentence description when one is provided, propose between two and six entities at the row level, together with their attributes. Entity names should be singular, and abbreviations should be expanded.} The model is not allowed to add example values to descriptions; these are appended later from the
database.

\item \textbf{Entity and attribute mapping}: \textit{Identify the table that corresponds to each proposed entity and the columns that correspond to its attributes.} If an attribute cannot be mapped with enough confidence, it is omitted rather than guessed.

\item \textbf{Table classification}: \textit{Classify each remaining table as an entity, link table, passthrough lookup, or technical table to ignore.} The prompt favors passthrough over ignore so that useful lookup data is not discarded. Deterministic checks may still reject a classification when the table's structure or lack of a valid identifier does not support it.

\item \textbf{Join verification}: \textit{Given the names, types, and sample values of two columns, along with the dataset description, decide whether their value overlap reflects a real relationship.} Candidates below the confidence threshold are rejected.

\item \textbf{Naming and refinement}: \textit{Generate short names for attributes reached through joins and revise generic labels using the values found in the column.} Name collisions trigger another model call, with a deterministic fallback if the conflict remains unresolved.

\item \textbf{Reference suggestion}: \textit{Propose a short label template for each entity using no more than two placeholders, with name-like attributes preferred.}

\end{itemize}

\end{document}